\documentstyle[prb,aps,epsf]{revtex}
\begin{document}
\draft
\title{Backward diode composed of a metallic 
 and  semiconducting nanotube}
\author{Ryo Tamura }
\address{Department of Physics, Graduate School of Science, University
  of Tokyo, Hongo 7-3-1, Bunkyo-ku, Tokyo 113, Japan}
\maketitle

\begin{abstract}
 The conditions necessary for a nanotube junction connecting
 a metallic and semiconducting nanotube 
 to rectify the current are theoretically investigated.
A tight binding model is used for the analysis,
 which includes  the Hartree-Fock approximation and the Green's function  method.
 It is found that the junction has a behavior  similar to the backward diode
 if the gate electrode is located nearby  and the Fermi level of the semiconducting tube is near the gap.
 Such a junction would be advantageous since
 the required length for the rectification could be reduced.
\end{abstract}
 
\newpage
\twocolumn

\newpage

%\section*{Introduction}
Carbon nanotubes (NTs)\cite{tube} are promising components for nano-electronics
 because they can either be  semiconducting (S) or  metallic (M) 
%\cite{tubeHamada,tubeSaito,tubeMint}.
\cite{tubeHamada}.
This has been explored in various experiments, for example, NT transistors \cite{Tans-transistor}, NT diodes \cite{Antnov-diode},  and NT junctions  
 \cite{junctionIijima,MSjunctionDekker}.
The NT junctions    made by a pentagon-heptagon defect pair 
 can be classified into three types: MM, 
MS, and SS: M and S stand for  a metallic NT and a semiconducting NT,
 respectively.
 It is remarkable that
 all the chemical bonds of NT junctions are
 essentially   $sp^2$ orbitals of the carbon with no impurity 
% \cite{junctionSaito,junctionChicoprl,junctionChicoprb,junctionDunlapldos,junctionMeunier,junctionCharlier},
 \cite{junctionSaito,junctionMeunier,junctionTamuraprb},
 which favors a coherent electron flow through the junction.
For MM-NT junctions, a scaling law of the electronic transport
 was found
%\cite{junctionTamuraprb,kpjunctionTamura,WFtamura,matsumura}.
\cite{junctionTamuraprb}.
On the other hand,  MS-NT junctions are expected to cause the Schottky barrier
 which is important to rectify the current \cite{odintsov}.
 In this work, we focus on MS-NT junctions connecting  the (9,0) tube and the (8,0) tube shown in Fig.\ref{3Dfig}.
 Though  in real experiments, 
these junctions are sometimes bent \cite{MSjunctionDekker},
 we investigate straight junctions because the difference does not  alter
 the results qualitatively
 as was explained in Ref.\cite{odintsov}.

For NT devices,
 the gate electrode is quite important for controlling the transport
 properties,  as was shown in an experiment where an
  AFM tip was used as the gate electrode 
%\cite{SPGM1,SPGM2}.
\cite{SPGM1}.
It was recently pointed out 
 that the length of the NT junction needs to be larger than the nanoscale
 in order to rectify the current because of the large depletion width in the underdoped case \cite{odintsov,Tersoff}.
 An important prospect is to understand whether or not the depletion
 width can be made narrower by approaching the gate closer to  the NT junction.
In this paper, we demonstrate that this could indeed be possible because  
 the gate electrode screens the Coulomb interaction
 in the junction.
 The main purpose of this paper is to support the validity of this method.
 
 The current  flowing through MS-NT junctions can be quite sensitive
 to the details of the Schottky barrier caused by the Coulomb interaction.
Though MS-NT junctions have been recently investigated theoretically \cite{odintsov,ferreira}, the Coulomb interaction
  was not considered in Ref.\cite{ferreira} and
 the semi-classical model used in Ref.\cite{odintsov} is not 
 appropriate when the spatial range of the 
 potential variation is short.
For these reasons we have adopted a tight binding model with 
 $\pi$ orbitals and the Hartree-Fock  approximation 
% \cite{harigaya,Junma,farajian}.
\cite{harigaya}.
 The corresponding Hamiltonian $H$ is represented 
 with the density matrix $\rho_{i,j}$ by
\begin{equation}
H_{i,j}=t_{i,j}-\frac{1}{2}U(|\vec{r}_i-\vec{r}_j|)
\rho_{i,j}
+\delta_{i,j}\sum_{k}U(|\vec{r}_i-\vec{r}_k|)
(\rho_{k,k}-1) \;\;,
\label{Hamil}
\end{equation}
 where $\vec{r}_j$ is the position  of the carbon atom $j$
  which can be determined approximately
 by the condition that all bonds must essentially 
 have the same length.
 This is justified since 
 the detailed atomic structure do not modify significantly the electronic
 structure \cite{junctionMeunier}.
The first term in Eq. (\ref{Hamil}) has a nonzero constant value $ -t < 0$ when $i$ and $j$ are nearest neighbors and is zero otherwise.
The second  and third term are the exchange and Hartree term,
 respectively.
The Coulomb interaction $U(r)$ is assumed to be $U(r)=t/\sqrt{1+4(r/a)^2}$ 
 where $a$ is the bond length.

The Green's function method 
 is used for calculating the transmission rate \cite{dattabook,tomanek}.
%\cite{tomanek,nardelli,anantramDatta,mehrez}.
The system is divided into three regions: the region near the defects ($C$), the (8,0) tube far from the defects ($S$), and the (9,0) tube far from the defects ($M$). The corresponding Hamiltonian $H$ is written as $H=H_C+H_S+H_M
+\tilde{t}_S+\tilde{t}_M$,
 where  $\tilde{t}_p$ ( $p=S$ or  $M$) represents the off-diagonal elements between the $C$
 region and the $p$ region.
For each region $p$, 
 the matrix element $(H_p+\tilde{t}_p)_{i,j}$  
 is written as $\tau_p$ when $i$ and $j$ are nearest neighbors, 
  $\epsilon_p$ when $i=j$, and zero otherwise.
 For each iteration step,
 the parameter $\epsilon_p$ and $\tau_p$
 are determined by the condition that the near 
 and far region are connected smoothly, i.e.
  $\epsilon_p$ and $\tau_p$ are taken to be the same as the corresponding
 averaged value of $(H_C)_{i,j}$ with the  atom $j$ directly connected to
  the region $p$.
The retarded Green's function $G(E)=(E+i\delta-H)^{-1}$ can then
 be represented by
\begin{equation}
 G_{k,l}= (E-H_C-\Lambda_S-\Lambda_M )^{-1}_{k,l}\;\;,
\label{green}
\end{equation}
 where  $k$ and $l$ belong to the region $C$.
 The effect of  the  region $S$ and  $M$ 
 are included in the `self energies', $\Lambda_p \equiv\tilde{t}_p(E+i\delta-H_p)^{-1}\tilde{t}_p$.
From Eq.(\ref{green}), the density matrix $\rho$ 
 in the region $C$  can be calculated
 as
\begin{equation}
\rho=\frac{1}{\pi}\int_{E_{F1}}^{E_{F2}}G\Gamma_{2}G^{*} dE
+\frac{1}{\pi}\int^{E_{F1}}_{-\infty}G(\Gamma_2+\Gamma_1)G^{*} dE\;\;.
\label{rho}
\end{equation}
Here $\Gamma_p \equiv  i(\Lambda_p-\Lambda_p^*)$,
$E_{F1}\equiv {\rm min}(E_{FM},E_{FS})$ and $E_{F2}\equiv{\rm max}(E_{FM},E_{FS})$. 
Thus, $H$ and $\rho$ can be calculated 
 self-consistently with Eqs.(\ref{Hamil}), (\ref{green}), and (\ref{rho}).
It is worth noting that the integrand of the second term in Eq.(\ref{rho})
 equals the DOS because $G(\Gamma_S+\Gamma_M)G^* =i(G-G^*)$ \cite{dattabook}.

Although the electron density $\rho_{i,i}$ in Eq.(\ref{rho}) is only defined
 for the region $C$, the Hartree terms caused by the charges outside this region
 should be included in $H_C$.
This can be done by repeating periodically
 that NT unit cell   in the region $C$ which is
  nearest to the far region.
Furthermore the electrostatic potential caused by the gate electrode
 can be obtained by the image charge method.
 Here we can choose the electrostatic potential to be zero at the gate surface
 without loss of generality
 because the difference, not the absolute value, of the electrostatic potential
 is relevant quantity.
Note that the nanotubes are not neutral by themselves, 
but the total system including the image charge {\it is} neutral.
In our model, the gate is planar and parallel to the tube axis.
Its distance from the tube axis 
 is chosen to be  three times  the radius 
 of the (9,0) tube.
 This is much smaller than the gate separation in Ref.\cite{odintsov}.
At this distance, we have verified that the orientation
 of the defects 
 does not change our main results.
In the following, the defects are assumed to face the gate (see Fig.\ref{3Dfig}).

When self-consistency has been achieved,
the current $I$ is calculated by 
\begin{equation}
 I=  (2e/h) \int_{E_{FM}}^{E_{FS}}{\rm Tr}[\Gamma_SG\Gamma_MG^{*}]  dE \;\;.
\label{current}
  \end{equation}
 The integrand of Eq.(\ref{current}) equals
  the total transmission rate $T(E) \equiv \sum_{i,j}T_{i,j}(E)$ where
  $T_{i,j}$ is the transmission  rate from the $i$th channel to the 
 $j$th channel. 
In the energy range considered here, both  tubes have 
 two channels except for the gap region of the (8,0) tube.
Note that $T(E)$ may be larger than one, 
 though $T_{i,j} \leq 1$.

%\section*{Result}
 The electron potential at each atom site 
  is shown by circles in Fig.\ref{band1}.
 The Fermi levels $(E_{FM},E_{FS})$ 
 shown by straight broken lines are
 $(-0.230t,-0.344t)$, $(-0.480t,-0.344t)$, $(-0.370t,-0.480t)$
 and  $(-0.600t,-0.480t)$  in (a), (b),  (c) and (d), respectively.
 These four situations are referred to(a),(b),(c) and (d),  hereafter.
The horizontal axis is the position along the tube axis and
 its range is as same as  the length of the region $C$.
In Fig.\ref{band1},
 the local gap energy region (hatched area)
 and the valence band edge (thin curved broken line) in the (8,0) tube
 are determined by the averaged electron potential 
 and the gap width of $H_S$.
 The Fermi levels correspond to hole doping which
  is not due to dopant atoms, but caused by 
  the source, drain and gate electrodes.
  The regions occupied by the electrons are shown by gray area.
 Though the energy range between $E_{FS}$ and $E_{FM}$ is
 $partially$ occupied (this corresponds to the first term
 of  Eq.(\ref{rho})), it  is not shown in Fig.\ref{band1}  explicitly.
 One may notice that several circles   show a sharp increase or rapid drop
  near the interface in Fig.\ref{band1}.
 These abrupt changes are caused by an excess of the electron on the pentagon
 and a deficiency on the heptagon \cite{junctionMeunier,Polytamura}.

The origin of the band bending 
 seen in Fig.\ref{band1} (b) and (d)
  is explained qualitatively with the `shifted Fermi level',
 $\tilde{E}_{Fp} \equiv E_{Fp}-\epsilon_p$, as follows.
Using the energy integral of the DOS,
 the hole density in the far regions
can be approximated
 by $\rho_M^h \sim C_M|\tilde{E}_{FM}|$ and 
 $\rho_S^h \sim C_S\sqrt{\tilde{E}_{FS}^2-\Delta^2}$ with the half
 gap width $\Delta$. 
 Here $C_M \simeq C_S$ because the two tubes have similar radii.
Then approximate values of the ratio $\rho^h_S/\rho^h_M$ are 0.32 and 0.44
  for Fig.\ref{band1}(b) and  (d), respectively, so
 $\rho^h_S < \rho^h_M$  in the both situations.
The charge distribution  connecting 
 $\rho^h_M$ to $\rho^h_S$ causes the electron potential
 to decrease toward the metallic tube, and is responsible for the
 band bending.
 Comparing Fig.\ref{band1}(b) and (d),
 one can see that both the height and  the width of the barrier
 decrease as the hole density $\rho^h_S$ increases.
 Qualitatively, this relation is similar  to that found in Ref.\cite{odintsov}.

The inset of Fig.\ref{trans} shows the dependence of
  the current   $I$ on
 the bias voltage  $V_{sd} \equiv (E_{FM}-E_{FS})/|e|$.
 The six open triangles and the five closed circles represent
 the data  when  $E_{FS}=-0.344t$ and  $E_{FS}=-0.480t$, respectively.
The data corresponding to Fig.\ref{band1} are indicated by the arrows.
 The horizontal separation of the 
 triangles or the circles  is not regular
  because each Fermi level $E_{Fp}$ was calculated
 with a fixed shifted Fermi level $\tilde{E}_{Fp}$.
 This was done since  the constant $\tilde{E}_{Fp}$
 keeps the $\rho^h_p$ unchanged in the iteration
 so that it is advantageous to achieve self-consistency.

The open triangles in the inset of Fig.\ref{trans} indicate
 that the current with constant $E_{FS}=-0.344t$ is rectified, i.e.
 the resistance is much higher when $V_{sd}<0$ than when $V_{sd}>0$.
  For example,
 $V_{sd}/I \sim$  32 k$\Omega$ for the case (b) and $V_{sd}/I \sim$  1300 k$\Omega$ for the case (a).
A positive $V_{sd}$, however, corresponds to a {\it reverse}
 bias for the thermo-electron current in conventional MS junctions,
 i.e.  the barrier becomes higher  as  $V_{sd}$ increases
 as can be seen in Fig.\ref{band1}.
The mechanism of this {\it reversed}
 rectification is similar to that of the backward diode
 \cite{backward}, which can
 be understood by examining
  the total transmission rate $T(E)$ as a function
 of the energy (Fig.\ref{trans}).
 Since the necessary energy range
 to calculate $I$ with Eq.(\ref{current}) is between $E_{FM}$ and $E_{FS}$,
 lines in Fig.\ref{trans} are limited to this range.
 $T(E)$ in the case (b) is considerably larger
 for the whole bias window
 because  the spatial 
 thickness of the barrier is reduced due to the proximity of the  gate 
 (see Fig.\ref{band1} (b)).
This leads to a large current for  positive $V_{sd}$.
 If the barrier  had been much thicker as in Ref.\cite{odintsov}, the current
 would have been close to zero.
On the other hand, 
  $T(E)$ in the case (a)  becomes zero
 for most of the bias window since the energy gap of the (8,0) tube
 blocks the current as can be seen in  Fig.\ref{band1} (a).
This example makes it clear that the necessary condition for keeping
 the current small when $V_{sd}$ is negative
 is to fix $E_{FS}$ near the  energy gap.
 Since $E_{FS}=-0.344t$ is closer to  the energy gap
 than $E_{FS}=-0.480t$, the triangles show much smaller current 
 than the circles in the inset when $V_{sd}<0$.
 This means that a better rectification has been achieved in
 the former case.
 Here we should mention that $ -(E_{FS}+E_{FM})/2$ was fixed 
, i.e. $E_{FS}$ depends on $V_{sd}$, 
in each I-V curve in Ref.\cite{odintsov}.
 This is not suitable for the rectification presented here.

In the present paper, we have to use a relatively smaller Coulomb interaction
 $U(r)=t/\sqrt{1+4(r/a)^2}$ than  that estimated  in other papers 
 \cite{harigaya},  since  the long spatial range of $U(r)$ makes the convergence  difficult.
 $U(r)$ is probably larger in the actual experiments
 than that used here, but it is expected that
 values of  larger $U(r)$ favor our conclusion.
 In fact, it is well known that the screening length $W$
 is determined by the relation $W \propto 1/\sqrt{U(0)N_F}$ 
  where $N_F$ is the DOS at the Fermi level.
Though this relation itself might not hold for MS-NT junctions,
 a decrease of $W$ with a simultaneous increase of $U(0)$  will hold and  
 favor our conclusion.

 In summary, we have calculated the
 I-V characteristics of a typical MS-NT junction.
 It is found that approaching the gate electrode to the junction
 and fixing $E_{FS}$ near the energy gap
 could be  an effective method to overcome 
 the problem of the large depletion length\cite{odintsov,Tersoff}.
 The rectifying MS-NT junction 
 discussed here is analogous to the backward diode.
 Though density functional calculations   will ultimately
 be necessary, the results obtained here
 are worth  testing in experiments.

 The author wishes to thank Professor Masaru Tsukada for stimulating
 discussions. 
 He also acknowledges Michel Gauthier for reading the manuscript
 and Shousuke Nakanishi for his technical support on the computer.
 This work was supported in part by the Grant-in-Aid for Scientific Research on
 Priority Area ``Fullerenes and Nanotubes'' by the Ministry of
 Education, Science and Culture of Japan.

%\bibliographystyle{prsty}
%\bibliography{sch_bib}

\begin{figure} % fig 1
\caption{The junction connecting the (9,0) tube and the (8,0) tube.
 Above right: setup of the electronic circuit. }
\label{3Dfig}
\end{figure}

\begin{figure} 
\caption{ The electron potential energy at each atom site.
 The Fermi levels $(E_{FM},E_{FS})$ 
 shown by straight broken lines are
 $(-0.230t,-0.344t)$,  $(-0.480t,-0.344t)$,  $(-0.370t,-0.480t)$
 and  $(-0.600t,-0.480t)$  in (a), (b), (c) and (d), respectively.
The unit of the vertical axis is the
 transfer integral $t$.
The horizontal axis represents the position along the tube axis and
 its unit is the lattice constant of graphite, i.e. $\sqrt{3}$ times
 the bond length $a$.
 Its origin is the interface between the (9,0) tube and the (8,0)
 tube and its range is the same as that of the near region $C$.
The horizontal arrows indicate the direction of the electron's flow.
 The local gap energy region, the valence band edge, 
 and the region occupied by electrons are shown
 by the thin curved broken line,
 the hatched area and the gray area, respectively.}
\label{band1}
\end{figure}

\begin{figure}
\caption{ Total transmission rate as a function of energy in the bias window.
The unit of the horizontal axis is the transfer integral $t$.
 Inset:  I-V characteristics of the junction
 when $E_{FS}$ is fixed to $-0.344t$ (open triangles) and $-0.480t$ (closed 
 circles). The units of the horizontal  and  vertical axis
 are $2et/h \sim$ 0.21 mA  and $t/|e|\sim$ 2.7 V, where $e$ and $h$ are 
the electron charge  and Planck's constant, respectively.
Data corresponding to Fig.2 are  indicated by the arrows.}
\label{trans}
\end{figure}

\end{document}